\documentclass[manuscript]{aastex}

\shorttitle{Remarks on the methods of investigations of alignment of galaxies}
\shortauthors{God{\l}owski}
 
\begin{document}
 
\title{Remarks on the methods of investigations of alignment of galaxies}
 
\author{W{\l}odzimierz God{\l}owski}
 
\affil{Uniwersytet Opolski, Institute of Physics, ul.  Oleska  48,
45-052 Opole, Poland}
\email{godlowski@uni.opole.pl}

\begin{abstract}
In the 1975 Hawley and Peebles gave the proposal to use three statistical tests
for investigations of the galaxies orientation in the large structures. Nowadays,
it has been considered as the standard method of searching for galactic
alignments. In the present paper we analyzed the tests in details and proposed
a few improvements. Basing on the improvements, the new method of analysis
of the alignment of galaxies in clusters is proposed. The power of this method
is demonstrated on the sample of 247 Abell clusters with at least 100 objects in
each. The distributions of the position angles for galaxies in each cluster are
analyzed using statistical tests: $\chi^2$, Fourier, autocorrelation and 
Kolmogorow test. The mean value of analyzed statistics is compared with 
theoretical predictions as well as with results obtained from numerical 
simulations. We performed 1000 simulations of 247 fictious clusters, each with 
numbers of galaxies the same as the real ones. We found that orientations of 
galaxies in analyzed clusters are not random i.e. that there exists an alignment
of galaxies in rich Abell galaxy clusters.

\end{abstract}
\keywords{galaxies: clusters: general}
 
\section{Introduction}
 
The analysis of the orientation of galaxies' planes is regarded as
a standard test of galaxies formation scenarios
\citep{Peebles69,Zeldovich70,Sunyaew72,Doroshkevich73,Shandarin74,Dekel85,Wesson82,Silk83,Bower06}.
Studies of the galaxies' planes orientation were conducted as early as in
19th century \citep{Abbe1875}. The review of early methods and results of
investigation performed up to the Second World War can be found in
the article of \citet{Danver42}, today only of historical value. The
first postwar work, which has been cited to this day, is the treatise of \citet{Holmberg46},
who  compared the numbers of galaxies seen  face-on  and
edge-on,  discussed the observational effects related to  optical
measurements of size of galaxy axes, and proved that the observed
excess  of  edge-on  galaxies is just  of  observational  origin.
 
In the early period after the Second World War the researchers were usually 
investigating distributions of position angles  within the galaxy-rich
regions (Cetus, Pisces, Hydra, Sextant, Ursa Maior, Virgo and Eridanus)
of the sky \citep{WB55,Br64,Br68}. In his two papers Brown \citep{Br64,Br68}
discovered  a departure from  isotropy in the distributions of position angles.
By analyzing the  distributions of position angles of large semiaxes of
galaxies,  \citet{Reinhardt70} and \citet{Reinhardt72} and later
\citet{Nilson74} found a very weak  preference  of  galaxy plane
alignment with the equator plane of the Local Supergalaxy. However
these results are undermined by the presence of background objects.
 
     Further important progress in the investigation of galaxies planes
orientation  was  made by \citet{h4}.  They discussed in  a detailed  manner
the method  of  investigating  the galaxies' orientation through analyzing 
distribution of position angles  as well  as the influence of possible errors
and observational effects. In particular, on the earlier papers of  Brown,
they indicated their insufficient certainty of his results due to possible 
errors in observations.
 
\citet{h4} analyzed the distributions of  position  angles using $\chi^2$
test, Fourier tests  and autocorrelation test. Since \citet{h4} this method
was accepted as standard method for analysis of an galactic alignment
\citep{Thompson76,MG82,Djorgovski83,f4,Kindl87,Flin88,f5,f6,Kampen90,Cerne90,g2,g3,Hu95,g4,Wu97,g5,Ar00,Baier03,Ar04,g05,Ar05a,Ar05b,Ar05c,Ar06a,Ar06b,Hu06,Wu97,Wu06,Ar07,Ar08,g10,g10a,Ar10,Ar11}
One should note that there are several modifications and improvements of
original \citet{h4} methods \citep{f4,Kindl87,g2,g3,g4,Ar00,g10a}. The aim of
the present paper is to present deeper improvements of the original \citet{h4}
method and show their usefulness for analysis of galactic orientations in
clusters. The power of this method is shown on the sample of 247 rich Abell
clusters.

Following \citet{g05} suggestion that alignment should increase with richness
of the cluster, it was found in \citet{g10a} that in rich Abell clusters
the non-randomness of the galaxies' orientation increased with number of objects
in clusters. The question which arose, is if we could say that
in analyzed sample of 247 Abell clusters with at least 100  objects  each,
we found an alignment. For this reason, in the present paper, we analyze
the distributions of position angles of galaxies belonging to investigated
clusters using $\chi^2$ test, Fourier  tests  and autocorrelation test applied
by \citet{h4} (see also \citet{f4,g2,g3,g10,g10a}) as well as Kolmogorow test.
For our sample of 247 Abell clusters, we compute the mean values of analyzed
statistics. Our null hypothesis $H_0$ is that the mean value of the analyzed
statistics is as expected in the cases of a random distribution of analyzed
angles. We compared our results with theoretical predictions as well as with
results obtained from numerical simulations.
 
\section{Observational data}
 
Our observational basis is the same as in \citet{g10a}. It is the sample
of 247 Abell clusters with at least 100 objects each, taken from PF catalogue
\citep{Panko06}. The structures were extracted from the Muenster Red Sky
Survey (MRSS hereafter) \citep{MRSS03}. MRSS is an optical large scale survey
covering the area of 5000 square degrees in the southern hemisphere with
$b < -45^o$. After scanning  217 ESO plates, it gives the information about
5,5 million galaxies. The 2D Voronoi tessellation technique was applied to
the MRSS galaxy catalogue to search for overdense regions \citep{Panko09}.
PF catalogue, like MRSS, is statistically complete till magnitude value
$m=18^m.3$ and it contains structures having at least ten members between
magnitude range $m_3$ and $m_3+3$ in each structure field. The $m_3$ is the
magnitude of the third brightest galaxy located in the considered structure
region. The resulting PF includes 6188 such structures. We select sample
of rich clusters (at least 100 members) being identified with one of ACO
clusters \citep{ACO}. There are 239 such objects in the PF catalogue.
Moreover, we include 9 objects which can be identified with two ACO clusters,
which increase our sample to 248 objects. However, we exclude from our
analysis A3822, which potentially has substructures \citep{Biviano97,Biviano02}.
Therefore, our sample has 247 objects.
 
The data for each galaxy member is taken from the MRSS. These includes:
the equatorial coordinates of galaxies($\alpha$, $\delta$), the diameters
of major and minor axes of the galaxy image ($a$ and $b$ respectively)
and the position angle of the major axis, $p$. Position angles are recomputed
from MRSS clockwise system to standard counterclockwise system.
We perform our computation both in Equatorial and Supergalactic Coordinate
System \citep{f4}. Because position angles for the face-on galaxies give
only marginal information connected with orientation of galaxy, we exclude
from analysis all galaxies with axial ratio $q=b/a>0.75$.
 
\section{The method of investigation}
 
The aim of our paper is to check if orientations of galaxies in investigated
clusters are isotropic. In order to check it we test if the distribution of
galaxy position angles $p$ (or supergalactic position angles $P$) is isotropic.
We apply statistical tests originally introduced  by \citet{h4} and later
modified by us \citet{g2,g3,g10a}, as well as Kolmogorow test.
In all considered tests,  the entire range of the tested $\theta$ angle
(where for $\theta$ one can put $p$ or $P$ respectively) is divided
into $n$ bins of equal width.  In the present paper we use $n = 36$.
 
Let $N$ denote the total number of galaxies in the considered cluster, and
$N_k$ - the number of galaxies with orientations within the $k$-th  angular
bin. Moreover, $N_{0,k}$ denotes expected number of galaxies in the $k$-th
bin. In our case all $N_{0,k}$ are equal $N_0$, which is also mean
number of galaxies per bin.
 
Our first test is $\chi^2$test:
\begin{equation}
\label{eq:c1}
\chi^2 = \sum_{k = 1}^n {(N_k -N\,p_k)^2 \over N\,p_k}= \sum_{k = 1}^n {(N_k -N_{0,k})^2 \over N_{0,k}}.
\end{equation}
where $p_k$ is a probability that chosen galaxy falls into $k$th bin.
We divided entire range of a $\theta$ angle into $n$ bins, which gives in
the $\chi^2$ test $(n-1)$ degrees of freedom. It means that expected
value $E(\chi^2)=n-1$ while variance $\sigma^2(\chi^2)=2(n-1)$.
For $n=36$ it gives $E(\chi^2)=35$ and variance $\sigma^2(\chi^2)=70$.
 
We analyzed sample of $m=247$ clusters. So we compute the mean value of
analyzed statistics i.e. in discussed case $\chi^2$ value, for whole sample
of clusters. If we assume uniform distribution of a $\theta$ angle, than
expected value $E(\bar{\chi^2})$ is again equal 35 while variance
$\sigma^2(\bar{\chi^2})= {\sigma^2(\chi^2) \over m}=0.2834$. It gives in our 
case standard deviation $\sigma(\bar{\chi^2})=0.5324$. We check our 
theoretical prediction by numerical simulations. We do this in two ways.
 
In the first case we simulate 247 fictious clusters, each with 2360 random
oriented members galaxies  and compute mean value of analyzed statistics
(i.e. in the discussed case $\chi^2$ value). We give 1000 simulations and on
this base we obtain: Cumulative Distribution Function (CDF) and Probability
Density Function (PDF). Expected value of analyzed statistics and their
variance is computed as well. Comparing them with theoretical prediction we
are able to check our theoretical assumptions, correctness of the program and
test quality of used random generator.
 
However, please note that the number of galaxies in our real clusters
is small in some cases, and the $\chi^2$ test will not necessarily
work well (e.g. the $\chi^2$ test requires the expected number of data
per bin to equal at least 7; see, however, \citet{SC67,Dom79}.)
For this reason we repeat this procedure but now we simulate 247 fictious
cluster each with number of members galaxies the same as in real clusters.
As a check, we repeat the derivations for different values of $n$, but
no significant difference appear so it is not presented in present paper.
 
Now we can compute the mean value of analyzed statistic for real sample
of analyzed 247 rich Abell clusters and compare it with theoretical
predictions and numerical simulations. This procedure is also provided for
other, presented below, analyzed statistics.

The first auto-correlation test quantifies the correlations between
galaxy numbers in neighboring angle bins. The measure of the correlation
is defined as
 
\begin{equation}
\label{eq:c2}
C\, = \, \sum_{k = 1}^n { (N_k -N_{0,k})(N_{k+1} -N_{0,k+1} )
\ \over \left[ N_{0,k} N_{0,k+1}\right]^{1/2} }
\end{equation}
where $N_{n+1}=N_1$. \citet{h4} noted that in the case of an
isotropic distribution, we expect $C = 0$ with the standard
deviation:
\begin{equation}
\label{eq:c3}
\sigma(C) = n^{1/2}
\end{equation}
 
Below we show that this result is an approximation which is not valid
in our case.
 
\citet{h4} result was obtained on the assumption that all $N_k$ are
independent from each other. One should note that $E(\Sigma\,X)=\Sigma\,E(X)$
and moreover, if two variables $X$ and $Y$ are independent than we have:
$E(X\,Y)=E(X)E(Y)$, $D^2(X+Y)=D^2(X)+D^2(Y)$ and
$D^2(X\,Y)=D^2(X)\,D^2(Y)+(E(X))^2\,D^2(Y)+D^2(X)\,(E(Y))^2$.
So, if all $N_k$ are independent than we obtain:

\begin{eqnarray}
\label{eq:c4}
&&E(C)\, = E\left(\, \sum_{k = 1}^n { (N_k -N_{0,k})(N_{k+1} -N_{0,k+1})
\ \over \left[ N_{0,k} N_{0,k+1}\right]^{1/2} } \right)=
\, \sum_{k = 1}^n{\,E\left( (N_k -N_{0,k})(N_{k+1} -N_{0,k+1})
\ \over \left[ N_{0,k} N_{0,k+1}\right]^{1/2} \right)}= \nonumber \\
&&=\, \sum_{k = 1}^n{\,E\left(N_k -N_{0,k}\over N_{0,k}^{1/2} \right)\,E\left(N_{k+1}-N_{0,k+1}\over N_{0,k+1}^{1/2}\right)}=0
\end{eqnarray}
and
\begin{eqnarray}
\label{eq:c5}
&&D^2(C)\, = D^2\left(\, \sum_{k = 1}^n { (N_k -N_{0,k})(N_{k+1} -N_{0,k+1} )
\ \over \left[ N_{0,k} N_{0,k+1}\right]^{1/2} }\right)=
\, \sum_{k = 1}^n\,D^2\left({ (N_k -N_{0,k})(N_{k+1} -N_{0,k+1} )
\ \over \left[ N_{0,k} N_{0,k+1}\right]^{1/2} }\right)=  \nonumber \\
&&=\, \sum_{k = 1}^n\,{D^2\left(N_k -N_{0,k}\over N_{0,k}^{1/2} \right)\,D^2\left(N_{k+1}-N_{0,k+1}\over N_{0,k+1}^{1/2}\right)}=
\, \sum_{k = 1}^n { 1 } =n
\end{eqnarray}
 
One should note however that the distribution of $N_k$ is in fact a polynomial
distribution and then elements of the covariance matrix of particular $N_k$
are given by the formulae $c_{ij}=N\,p_i(\delta_{ij}-p_j)$. If two variables
$X$ and $Y$ are not independent than $E(X\,Y)=E(X)E(Y)+Cov(X\,Y)$.
It leads to a conclusion that now in our formulae for $E(C)$ is present an
additional term connected with covariance between value of $N_k$ and $N_{k+1}$.
As a result $E(C)\, =-\sum_{k = 1}^n{\,N\,p_k\,p_{k+1}}$. Because in our case
all $p_k$ (and as a result also $N_{k,0}$) are equal ($p_k=1/n$) than
$E(C)\, =-\sum_{k = 1}^n\,{p_k}=-\sum_{k = 1}^n\,{1/n}=-1$. Moreover when
we tray to compute $D^2(C)$ than variance of $C$ contain term which is
variance of products of $N_k$ and $N_{k+1}$ which are not independent. So,
correct value of $D^2(C)$ is different from $n$ and is obtained from numerical
simulations.
 
Differences between our result and \citet{h4} approximation is not significant
in the case of individual clusters because the difference between results in
expected value of $C$ (0 or -1) is small with comparison to its standard
deviation $\sigma(C) \approx \sqrt{n}$. However one should note that in our
case this difference is important. Our sample has 247 clusters so standard
deviation of $\bar{C}$, $\sigma(\bar{C}) \approx \sqrt{n/247} = 0.3818$ is
significantly smaller than a difference in expected values (which is equal to $1$).
 
If deviation from isotropy is a slowly varying function of the angle $\theta$
one can use the Fourier test \citep{h4}
 
\begin{equation}
\label{eq:f1}
N_k = N_{0,k} (1+\Delta_{11} \cos{2 \theta_k} +\Delta_{21} \sin{2\theta_k})
\end{equation}
 
We obtain the following expression for the $\Delta_{i1}$ coefficients
 
\begin{equation}
\label{eq:f2}
\Delta_{11} = {\sum_{k = 1}^n (N_k -N_{0,k})\cos{2 \theta_k} \over
\sum_{k = 1}^n N_{0,k} \cos^2{2 \theta_k}},
\end{equation}
 
\begin{equation}
\label{eq:f3}
\Delta_{21} = { \sum_{k = 1}^n (N_k-N_{0,k})\sin{2 \theta_k} \over
\sum_{k = 1}^n N_{0,k} \sin^2{2 \theta_k}}.
\end{equation}
 
These equations are originaly introduced by \citet{h4}. It was written in
a simple explicite form  in the case when all $N_k$ was equal and and n=36
(equation 25 \citet{h4}).

Standard deviation of $\sigma(\Delta_{11})$ and $\sigma(\Delta_{12})$ is given
by expressions:
\begin{equation}
\label{eq:f4}
\sigma(\Delta_{11}) =
 \left( {\sum_{k = 1}^n N_{0,k} \cos^2{2 \theta_k} } \right)^{-1/2} =
 \left( {2 \over n N_0} \right)^{1/2},
\end{equation}
\begin{equation}
\label{eq:f5}
\sigma(\Delta_{21}) =
 \left( {\sum_{k = 1}^n N_{0,k} \sin^2{2 \theta_k}} \right)^{-1/2} =
 \left( {2 \over n N_0} \right)^{1/2}.
\end{equation}
 
The probability that the amplitude
\begin{equation}
\label{eq:f6}
\Delta_1 = \left( \Delta_{11}^2 + \Delta_{21}^2 \right)^{1/2}
\end{equation}
is greater than a certain chosen value is given by the formula
\begin{equation}
\label{eq:f7}
P(>\Delta_1 ) = \exp{\left( -{n \over 4} N_0 \Delta_1^2 \right)}
\end{equation}
with standard deviation of this amplitude
\begin{equation}
\label{eq:f8}
\sigma(\Delta_1) = \left( {2 \over n N_0} \right)^{1/2}.
\end{equation}
 
The formula for standard deviation for $\sigma(\Delta_{11})$,
$\sigma(\Delta_{12})$, and $\sigma(\Delta_{1})$ ($\Delta$ in the original
\citet{h4} notation) was also written in a simple explicit form in
\citet{h4} (equation 26). This test was substantially improved by
\citet{g3} for the case when higher Fourier mode is taken into account:
\footnote{However please note that there is a printed error in
God{\l}owski (1994). Eq. 18 should have form $P(\Delta)=(1+J/2)\exp{(-J/2)}$}
 
\begin{equation}
\label{eq:f9}
N_k = N_{0,k} (1+\Delta_{11} \cos{2 \theta_k} +\Delta_{21} \sin{2
\theta_k}+\Delta_{12} \cos{4 \theta_k}+\Delta_{22} \sin{4\theta_k}+.....).
\end{equation}
 
In our case (all $N_{0,k}$ are equal) it leads to formulas for the
 $\Delta_{ij}$ coefficients \citep{g10a}:
\begin{equation}
\label{eq:f10}
\Delta_{1j} = {\sum_{k = 1}^n N_k\cos{2J \theta_k} \over
\sum_{k = 1}^n N_{0} \cos^2{2J \theta_k}},
\end{equation}
and
\begin{equation}
\label{eq:f11}
\Delta_{2j} = { \sum_{k = 1}^n N_k\sin{2J \theta_k} \over
\sum_{k = 1}^n N_{0} \sin^2{2J \theta_k}},
\end{equation}
with the standard deviation
\begin{equation}
\label{eq:f12}
\sigma(\Delta_{1j}) =
\left( {\sum_{k = 1}^n N_{0} \cos^2{2J \theta_k} } \right)^{-1/2} =
 \left( {2 \over n N_0} \right)^{1/2},
\end{equation}
and
\begin{equation}
\label{eq:f13}
\sigma(\Delta_{2j}) =
 \left( {\sum_{k = 1}^n N_{0} \sin^2{2J \theta_k} } \right)^{-1/2} =
 \left( {2 \over n N_0} \right)^{1/2}.
\end{equation}
 
If we analyze Fourier modes separately, probability that the amplitude
\begin{equation}
\label{eq:f14}
\Delta_j = \left( \Delta_{1j}^2 + \Delta_{2j}^2 \right)^{1/2}
\end{equation}
is greater than a certain chosen value is given by the formula:
\begin{equation}
\label{eq:f15}
P(>\Delta_j ) = \exp{\left( -{n \over 4} N_0 \Delta_j^2 \right)}.
\end{equation}
 
When we analyze first and second Fourier modes together the probability that
the amplitude
\begin{equation}
\label{eq:f16}
\Delta =
 \left( \Delta_{11}^2 + \Delta_{21}^2+\Delta_{12}^2 + \Delta_{22}^2 \right)^{1/2}
\end{equation}
is greater than a certain chosen value is given by the formulae
\begin{equation}
\label{eq:f17}
P(>\Delta ) =
\left(1+{n \over 4} N_0 \Delta_j^2 \right) \exp{\left( -{n \over 4} N_0 \Delta_j^2 \right)}.
\end{equation}
 
The value of coefficient $\Delta_{11}$ gives us the direction of departure
from isotropy. If, $\Delta_{11}<0$ then the excess of the galaxies with
position angles near $90$ degrees is observed. It means that in this case
excess of galaxies with the position angel parallel to equatorial plane
(case $p$) or parallel to Local Supercluster plane (case $P$) is observed,
while for $\Delta_{11}>0$ the  excess of the galaxies with the position angle,
respectively perpendicular to equatorial plane or Local Supercluster plane
is observed.

In the paper \citet{g10a} the investigation of the linear regression given
by  $y=aN+b$ counted for various parameters was performed. In the case of
position angles the linear regression between the values of statistics
$\chi^2$, $\Delta_1/\sigma(\Delta_1)$, $\Delta/\sigma(\Delta)$ and the number
of analyzed galaxies in  each particular cluster was studied. It was found
that non-randomness of galaxy orientation increased with numbers of objects
in clusters. To test our null hypothesis $H_0$ that value of the
analyzed statistics is as expected in the cases of random distribution of
analyzed angles, we should now discuss properties of statistics
$\Delta_1/\sigma(\Delta_1)$, $\Delta/\sigma(\Delta)$ as well as properties
of the whole Fourier test in more details.
 
In one dimensional ($1D$) case the situation is very clear. Variables
$\Delta_{11}/\sigma(\Delta_{11})$, $\Delta_{21}/\sigma(\Delta_{21})$
are normalized gausian variables i.e. with expected value equal $0$ i.e.
$E(\Delta_{ij}/\sigma(\Delta_{ij}))$=0 and Variance equal $1$ i.e.
$D^2(\Delta_{ij}/\sigma(\Delta_{ij}))=1$. Of course
$\sigma^2(\overline{\Delta_{ij}/\sigma(\Delta_{ij})})=1/247=0.00405$ and
$\sigma (\overline{\Delta_{ij}/\sigma(\Delta_{ij})})=0.06363$. However, in
the case of $\Delta_1/\sigma(\Delta_1)$ and $\Delta/\sigma(\Delta)$ variables
situation is much more complicated.
 
In our case, taking into account the equation 26 \citet{h4} (our equation
\ref{eq:f8}) the equations \ref{eq:f7} and \ref{eq:f15} can be written
(in analogy to $1D$ gaussian distribution) in the form:
\begin{equation}
\label{eq:f18}
P(>\Delta_j ) = \exp{\left( -{1 \over 2} {\Delta^2_j \over \sigma^2(\Delta_j)} \right)}.
\end{equation}
while (having in mind that $\Delta$ is given by the equation \ref{eq:f16})
the equation \ref{eq:f17} could be  written as:
\begin{equation}
\label{eq:f19}
P(>\Delta) =
\left(1+{1 \over 2}  {\Delta^2 \over \sigma^2(\Delta)} \right) \exp{\left( -{1 \over 2} {\Delta^2 \over \sigma^2(\Delta)}  \right)}.
\end{equation}
 
Please note however that $\Delta_j$ is described by $2D$ Gaussian
distribution while $\Delta$ is described by $4D$ Gausian distribution.
In an explicit form the equation \ref{eq:f7} could be written as:
\begin{equation}
\label{eq:f20}
P(>\Delta_1 ) =
\exp{\left( -{1 \over 2} \left( {\Delta^2_{11} \over \sigma^2(\Delta_{11})}+{\Delta^2_{21} \over \sigma^2(\Delta_{21})} \right)\right)}.
\end{equation}
So, the notation $\Delta^2_j \over \sigma^2(\Delta_j)$ means only that
elements of $\Delta^2$ should be devided by elements
of covariance matrix $\Delta_{ij}$. Even more generally it could be
written as:
 
\begin{equation}
\label{eq:f21}
P(>\Delta_1 ) = \exp{\left( -{1 \over 2} \sum_i \sum_j {G_{ij} I_{i} I_{j}} \right)}.
\end{equation}
where $I$ vector is
\begin{equation}
\label{eq:f21a}
I=
\left(
\begin{array}{c}
\Delta_{11} \\
\Delta_{21}
\end{array}
\right)
\end{equation}
and the matrix G is the inverse matrix to the covariance matrix of
$\Delta_{ij}$ ($Cov=G^{-1}$). In the $4D$ case vector $I$ has a form:
\begin{equation}
\label{eq:f21b}
I=
\left(
\begin{array}{c}
\Delta_{11} \\
\Delta_{21} \\
\Delta_{12} \\
\Delta_{22}
\end{array}
\right)
\end{equation}
while (using an auxiliary variable $J=\sum_i \sum_j {G_{ij} I_{i} I_{j}}$)
\begin{equation}
\label{eq:f21c}
P(>\Delta)=(1+J/2)\exp{(-J/2)}
\end{equation}
Please note that the equation \ref{eq:f21} is $2D$ equivalent of the equation
\ref{eq:f21c} (and the equation 18 of \citet{g3} paper). Please note that the
advantage of our notation is that it could be very easily extended on the
situation that not all $\sigma(\Delta_{ij})$ are equal and/or non diagonal
elements of covariance matrix are not disappear (i.e. not all $\Delta_{ij}$
are independent to each other).
 
One should note that the equation \ref{eq:f8} (\citet{h4} equation 26) is
obtained as a result of the theorem of propagation of errors. Because in
our case $\sigma(\Delta_{11})=\sigma(\Delta_{21})=(2/N)^{1/2}$ and
$\Delta_1 = \left( \Delta_{11}^2 + \Delta_{21}^2 \right)^{1/2}$
we obtain the following results:
\begin{eqnarray}
\label{eq:f22}
&&\sigma^2(\Delta_1) =
\left(\partial\Delta \over {\partial\Delta_{11}} \right)^2 \sigma^2(\Delta_{11})+
\left(\partial\Delta \over {\partial\Delta_{21}} \right)^2 \sigma^2(\Delta_{21})= \nonumber \\
&& = \left(2\Delta_{11} \over {2\sqrt{\Delta^2_{11}+\Delta^2_{21}}}\right)^2\sigma^2(\Delta_{11})+
\left(2\Delta_{21} \over {2\sqrt{\Delta^2_{11}+\Delta^2_{21}}}\right)^2\sigma^2(\Delta_{21})= \nonumber \\
&& = \sigma^2(\Delta_{11})=\sigma^2(\Delta_{21})=2/N
\end{eqnarray}
i.e. our equation \ref{eq:f8} (or \citet{h4} equation 26).
We obtain analogical results in the case when we are taking into account
both first and second Fourier modes
$\sigma^2(\Delta)=\sigma^2(\Delta_{11})=\sigma^2(\Delta_{21})=\sigma^2(\Delta_{12})=\sigma^2(\Delta_{22})=2/N$.
However, please note that the theorem of propagation errors is obtained
in the linear model. We argue below that a linear approximation is not good
approximation in our case.
 
Please note that a value beeing the sum $\Delta^2_{ij}/\sigma^2(\Delta_{ij})$
is $\chi^2$ distributed \citep{Brandt97}. As a result
the value
\begin{equation}
\label{eq:f23}
\Delta^2_{1}/\sigma^2(\Delta_{1})=\Delta^2_{11}/\sigma^2(\Delta_{11})+\Delta^2_{21}/\sigma^2(\Delta_{21})
\end{equation}
has  $\chi^2$ distribution with 2 degree of freedom while
\begin{equation}
\label{eq:f24}
\Delta^2/\sigma^2(\Delta)=\Delta^2_{11}/\sigma^2(\Delta_{11})+\Delta^2_{21}/\sigma^2(\Delta_{21})+
\Delta^2_{12}/\sigma^2(\Delta_{12})+\Delta^2_{22}/\sigma^2(\Delta_{22})
\end{equation}
has  $\chi^2$ distribution with 4 degree of freedom.
So, $E(\Delta^2_{1}/\sigma^2(\Delta_{1}))=2$, $E(\Delta^2/\sigma^2(\Delta))=4$,
$D^2(\Delta^2_{1}/\sigma^2(\Delta_{1})=4$ and $D^2(\Delta^2/\sigma^2(\Delta)=8$.
Because we analyzed the $\Delta_1/\sigma(\Delta_1)$ and $\Delta/\sigma(\Delta)$
statistics, the expected value and the standard deviation of this statistics
are of our interest. The theorem of propagation errors shows that
\begin{equation}
\label{eq:f25}
\sigma^2(x)=\left(\partial x \over {\partial x^2}\right)^2\sigma^2(x^2)=
\left(1 \over {2 \sqrt{x^2}} \right)^2 \sigma^2(x^2)= {\sigma^2(x^2) \over 4x^2}
\end{equation}
In our case it leads to results that
$\sigma(\Delta_{1}/\sigma(\Delta_{1}))=1/2$, and  again
$\sigma(\Delta/\sigma(\Delta))=1/2$.
The expected value of $E(X)=\sqrt{E(X^2)-D^2(X)}$. In our case
\begin{equation}
\label{eq:f26}
E\left({\Delta_1 \over \sigma(\Delta_1)}\right)=
\sqrt{E\left(\Delta^2_{1} \over \sigma^2(\Delta_{1})\right)-D^2\left(\Delta_1 \over \sigma(\Delta_1)\right)}=
\sqrt{2-0.5}=1.2247
\end{equation}
and
\begin{equation}
\label{eq:f27}
E(\left({\Delta \over \sigma(\Delta)}\right)=
\sqrt{E\left(\Delta^2 \over \sigma^2(\Delta )\right)-D^2\left(\Delta \over \sigma(\Delta)\right)}=
\sqrt{4-0.5}=1.8708
\end{equation}
Because in our sample we have 247 clusters than
$\sigma^2(\overline{\Delta_{1}/\sigma(\Delta_{1})})$ and
$\sigma^2(\overline{\Delta/\sigma(\Delta)})$ is equal
${1/2 \over 247}=0.002024$ while standard deviation of
$\sigma(\overline{\Delta_{1}/\sigma(\Delta_{1})})$ and
$\sigma(\overline{\Delta/\sigma(\Delta)})$ is equal
$\sqrt{1/2 \over 247}=0.04499$. These results are obtained from theorem of
propagation errors, so again we assume linear approximation. We show in the
next section that in presently analyzed case it works quite well, but
correct values must be obtained from  numerical simulations.
 
The isotropy of the resultant distributions of the angles $\theta$ can also
be investigated using Kolmogorov- Smirnov test (K-S test).
We assume that the theoretical, random distribution contains the same
number of objects as the observed one. In such a case statistics $\lambda$
\begin{equation}
\label{eq:k1}
\lambda=\sqrt{n}\,D_n
\end{equation}
is given by limit Kolmogorow distribution, where
\begin{equation}
\label{eq:k2}
D_n= sup|F(x)-S(x)|
\end{equation}
and F(x) and S(x) are theoretical and observational distributions of $\theta$.
Now we can compute mean value and standard deviation of analyzed statistic
for real sample of analyzed 247 rich Abell clusters.
Expected value of $\lambda$, its standard deviation as well as PDF and CDF
of $\lambda$ are obtained from numerical simulations.
 
\section{Numerical Simulations}
 
A well known problem with random number generators is that their quality
is difficult to asses in any rigorous way. In fact many of the popular
generators used till now failed to give correct results in multidimensional
(sometimes even in two dimensional) simulations \citep{Luescher94}.
We decided to test a few different number generators to be sure that our
result is correct. The first one is built in Fortran Lahey v3 generator.
Moreover we used classical RAN1 generator with versions ($Ran1_{pt}$) of 
Numerical Recipes \citep{Press85} and more recent version ($Ran1_{nt}$) of 
\citet{PT92}, as well as GGUBS subroutine from ISML library, "minimal standard 
generator" discussed by \citet{Park88}. We compared the result obtained
using above generators with a result obtained with our basic generator
which is a subtract-and-borrow random number generator originally proposed by
\citet{MZ91} and later improved by Martin Luescher and called RANLUX (level 4) 
generator \citep{Luescher94,James94}. 

The work of Martin Luescher provides the first operational definition of randomness 
in the sense required by Monte Carlo calculations and the Ranlux generator 
is the first one which produces random sequence in which no defect can be observed. 
The period of the generator is about $10^{171}$. The RANLUX generator is based on 
a dynamical system which may be regarded as a multi-dimensional version of Arnold's 
famous cat map. Similarly to the cat map, the system can be proved to be chaotic in 
a strong sense \citet{Luescher94,Luescher10}.

Fundamentally different between traditional random number generators (TRNG) 
and RANLUX is that TRNG's must be tested because, apart from the testing, there 
is no reason to believe they are at all random.  Experience shows that testing 
is necessary but not sufficient. RANLUX, on the other hand, has a good underlying 
theory, so the purpose of testing is only to make sure that the theory has been 
understood, applied and programmed correctly \citet{James94}. Discussion of 
different types of Random Generator and advantages of RANLUX was discussed for 
example by \citet{Shchur98}.
 
We performed  1000 simulations of 247 fictious clusters, each with 2360 random
oriented members galaxies. As a result we have the sample of $l=1000$ values of
particular statistics. For the present analysis we choose $\chi^2$, $C$ and
$\Delta_{11}/\sigma(\Delta_{11})$ statistics for which we have good theoretical
predictions only with an exception of the variance for $C$ statistics which is
obtained from the Peebles approximation. In this Table \ref{tab:t1} we present
average value of the analyzed statistics, its standard deviation and standard
deviation in the sample. Moreover we present the standard deviation of the
standard deviation estimator $S$ which is equal $\sigma(S)=S/\sqrt{2(l-1)}$
\citep{Brandt97}.
 
One should note that RAN1 generator does not survive the tests. Version
 of Numerical Recipes \citep{Press85} gives wrong result of $C$ 
statistics on more than $40\,\sigma$ level and wrong result of $\chi^2$ on 
more than $30\,\sigma$. \citet{PT92} version is better but gives wrong result 
of $\Delta_{11}/\sigma(\Delta_{11})$ statistics on more than $10\,\sigma$ 
level, and more than $2\sigma$ deviation for $\chi^2$ statistic. With built 
in Fortran Lahey v3 generator and GGUBS subroutine, a situation is much 
better, however some deviations from expected value (up to $2\sigma$ level) 
are observed. The RANLUX generator satisfies all three tests and we choose 
this generator as our base generator.
 
In our further analysis we choose six tests. We analyze $\chi^2$,
$\Delta_1/\sigma(\Delta_1)$, $\Delta/\sigma(\Delta)$, $C$, $\lambda$ and
$\Delta_{11}/\sigma(\Delta_{11})$ statistics. In the Table \ref{tab:t2} we
present (as in the Table \ref{tab:t1}) average value of the analyzed
statistics, its standard deviation, standard deviation in the sample as well
as its standard deviation. Because of small number of galaxies in some
cluster we repeat our analysis (Table \ref{tab:t3}) with 1000 simulations of
247 fictious clusters, each cluster with number of members galaxies the same
as in the real cluster. On the base of these simulations we built PDF and CDF
presented in the Figures {\ref{fig:f1} and {\ref{fig:f2}.

In our procedure we compute the mean values of the analyzed statistics. When 
the errors are normal (Gaussian), what is true at least in the case of 
$\Delta_{11}/\sigma(\Delta_{11})$ statistic, that parameters are estimated
by maximum-likelihood method. They should have asymptotic normal (Gausian)
distribution. Now we check this suppositions using K-S test. For the test we 
choose the statistics $\bar{\chi^2}$ and $\overline{\Delta_{11}/\sigma(\Delta_{11})}$
because we have well theoretical predictions about both expected values 
and variances of these statistics (one should note however that $\chi^2$ statistics 
is $\chi^2$ not normal distributed). As it was shown in the previous sections 
for $\bar{\chi^2}$ statistic $E(\bar{\chi^2})=35$ and $\sigma^2(\bar{\chi^2})=0.2834$
while for $\overline{\Delta_{11}/\sigma(\Delta_{11})}$ statistic 
$E(\overline{\Delta_{11}/\sigma(\Delta_{11})})=0$ and 
$\sigma^2(\overline{\Delta_{11}/\sigma(\Delta_{11})})=0.00405$.
In order to reject the $H_0$ hypothesis, that distribution is Gaussian with 
expected value and variance as above, the value of observed statistics $\lambda$ 
should be greater than $\lambda_{cr}$. At the significance level $\alpha = 0.05$
the value $\lambda_{cr}$ = 1.358. 

In the case when we performed  1000 simulations of 247 fictious clusters, each 
with 2360 random oriented members galaxies (Table \ref{tab:t2}) we obtained the values 
of statistic $\lambda$ equal $0.5443$ in the case of $\bar{\chi^2}$ statistic 
and $0.7963$ in the case of $\overline{\Delta_{11}/\sigma(\Delta_{11})}$ statistic.
When we repeated our analysis  with 1000 simulations of 247 fictious clusters, each 
cluster with number of members galaxies the same as in the real cluster 
(Table \ref{tab:t3}) and Figures {\ref{fig:f1} and {\ref{fig:f2}), we obtained 
values $\lambda=0.9591$ and $\lambda=0.7229$ respectively. All these valuese of 
$\lambda$ are significantly less than $\lambda_{cr} = 1.358$. So we can not exclude 
our $H_0$ hypothesis. One should note however, that values of $\bar{\chi^2}$ and 
$\overline{\Delta_{11}/\sigma(\Delta_{11})}$ statistics obtained from numerical
simulations ((Tables \ref{tab:t2} and \ref{tab:t3}) are a litle bit different from
theoretical one. 

\citet{Lilliefors67} showed that the standard tables used for the Kolmogorov-Smirnov 
test are valid when testing whether a set of observations are from a completely specified 
continuous distribution. When we check if the distribution is normal but one or more  
parameters are estimated from the sample then the Kolmogorov-Smirnov test no longer 
applies. At least it is not allowed to use the commonly tabulated critical points. It 
is suggested by \citet{Massey51} that if the test is used in this case, the results will 
be conservative in the sense that the probability of a type $I$ error will be smaller 
than as given by tables of the Kolmogorov-Smirnov statistic. \citet{Lilliefors67} 
showed that the results of this procedure will indeed be extremely conservative and 
computed a new table for critical value of $D = max |F^*(x) - S_N(x)|$ statistic, where 
$S_N(x)$ is the observational cumulative distribution function and $F^*(x)$ is the 
cumulative normal distribution function with mean value and variance estimated from 
the sample. His critical values are about $30\%$ less than those  obtained by 
\citet{Massey51} for classical Kolmogorov-Smirnov test. 

Above modification of  the Kolmogorow test is usualy known as Kolmogorow - Lilliefors 
test. For this test the critical value of $D_{cr}$, at the significance level 
$\alpha = 0.05$ for $n=1000$ is equal $0.028$. For analyzed statistics we obtained 
folowing values of $D$. For the sample of 247 fictious clusters, each with 2360 random 
oriented members galaxies (Table \ref{tab:t2}) we obtained $D=0.014$ in the case of
$\bar{\chi^2}$ statistic and $D=0.018$  in the case of 
$\overline{\Delta_{11}/\sigma(\Delta_{11})}$ statistic. For sample of 247 fictious 
clusters, each cluster with number of members galaxies the same as in the real 
cluster (Table \ref{tab:t3}) we obtained $D=0.018$ in the case of $\bar{\chi^2}$ 
statistic and $D=0.0135$ in the case of $\overline{\Delta_{11}/\sigma(\Delta_{11})}$ 
statistic. Again all these values of $\lambda$ are significantly less than 
critical value and  again we can not exclude our $H_0$ hypothesis.

As a result we can conclude that analyzed statistics can be well described by the 
normal distribution. In particular it means that fluctuations observed in the 
Figure {\ref{fig:f2} in PDF of $\bar{\chi^2}$ statistics are not in conflict with our 
prediction that the statistics is normally distributed with parameters as obtained 
from theoretical predictions.

Independently we analyze situation (Table \ref{tab:t4}) when only galaxies brighter 
than $m_3+3$ are taken into account. Comparison of the Tables \ref{tab:t2}, 
\ref{tab:t3} and \ref{tab:t4} shows that there are some but not big differences 
between particular cases. When we compare average values of the statistics in the 
Tables \ref{tab:t2} and \ref{tab:t3} we found that in all cases diferences beetwen 
them are less than theirs $3$ standard deviations, however for $\Delta_1/\sigma(\Delta_1)$, 
$C$ and $\lambda$ these diferences are on the $2\sigma(\bar{x})$ level. It confirmed 
our sugestion that PDF and CDF used in further analyzis should be built on the base 
of 1000 simulations of 247 fictious clusters, each cluster with number of members 
galaxies the same as in the real cluster. In the case of Tables \ref{tab:t3} and 
\ref{tab:t4} diferences between average values are on $2\sigma(\bar{x})$ level only 
in the case of $\Delta/\sigma(\Delta)$ test. So it is not necessary  to build special 
PDF and CDF for this case.

\section{Results}
 
We analyzed the distribution of the position angles in the sample (A) of 247
rich Abell clusters both in Equatorial and Supergalactic coordinate system.
Moreover we analyzed restricted sample (B) in which only galaxies brighter
than $m_3+3$ are taken into account. The results are presented in the Table
\ref{tab:t5}.
 
Our null hypothesis $H_0$ is that mean value of the analyzed statistics
is as expected in the cases of a random distribution of the position angles,
against $H_1$ hypothesis that analyzed values are different than in the case
of random distribution. For the $\chi^2$ test the result is significant on
$3\,\sigma$ level, for autoccorelation test it is significant on $4\,\sigma$
level, while for Fourier test ($\Delta_1/\sigma(\Delta_1)$ and
$\Delta/\sigma(\Delta)$ statistics) and Kolmogorow test ($\lambda$ statistics)
the results are significant on more than $5\,\sigma$ level. In all cases
there are no significant differences when we analyzed distribution
of Equatorial position angles $p$ and Supergalactic position angles $P$.
One can see from PDF and CDF presented in the Figures {\ref{fig:f1} and 
\ref{fig:f2} that probability that  such results are coming from random 
distributions is (in all cases) less than $0.1\%$.
 
$\Delta_{11}/\sigma(\Delta_{11})$ test does not show any difference from
predictions of our null hypothesis $H_0$ which means that the average value
of the analyzed statistic is as expected in the cases of a random
distribution. This result, together with the fact that we show no significant
differences with analysis in Equatorial and Supergalactic coordinate systems,
shows that observed alignment is not connected with equatorial plane
(as expected) nor with Supergalactic plane. Interpretation of this conclusion 
in the context of evolution of galaxies in the cluster needs detailed future 
investigation. 

In our opinion it is because of the influence of environmental effects to the 
origin of galaxy angular momenta. \citet{gpf11}  studied the galaxy alignment 
in the sample of very rich Abell clusters located in and outside superclusters. 
Even though that orientations of galaxies in analyzed clusters are not random,
both in the case when we analyzed whole sample of the clusters and only clusters 
belonging to the superclusters, the statistically significant difference among 
investigated samples was found. In contrast to whole sample of cluster, where 
alignment increases with the cluster richness \citet{g10a} the cluster belonging 
to the superclusters does not show this effect. Moreover, the alignment
decreases with the supercluster richness. The observed trend,dependence of 
galaxy alignment on both cluster location and supercluster richness clearly 
supports the influence of environmental effects to the origin of galaxy angular 
momenta.

Another important possibility is the influence of of the large scale orientation
of galaxy clusters \citep{h6,Wang09,g10,Paz11,VC11,Hao11,Sch11,S11,Blazek11,Song11,Noh11} 
analized both theorticaly and observationaly.
\citet{g10} studied  the orientation of galaxy groups in the Local Supercluster (LSC).
It is strongly correlated with the distribution of neighbouring groups in the scale 
till about 20 Mpc.
\citet{Paz11} found  a strong alignment between the projected major axis of group 
shapes and the surrounding galaxy distribution up to scales of $30 Mpc/h$. 
\citet{S11} search for two types of cluster alignments using pairs of clusters: 
the alignment between the projected major axes of the clusters founding weak effect 
up to $20 Mpc/h$, and the alignment between one cluster major axis and the line 
connecting it to the other cluster in the pair  founding strong  alignment on scales 
up to $100 Mpc/h$. 

The change of alignment with the surrounding neigbourhood
was observed also in alignment study in void vicinity \citep{v11}
being continuation of earlier study of galaxy orientation in
regions surrounding bubble-like voids \citep{t06}.
Another interesting result was found by \citet {Jones10} who reported that the 
spins of spiral galaxies located within cosmic web filaments tend to be aligned 
along the larger axis of the filament, which is interpreted by the authors as 
"fossil" evidence indicating that the action of large scale tidal torques effected 
the alignments of galaxies located in cosmic filaments.

For the sample B the results are weaker but still significant. As above,
$\Delta_{11}/\sigma(\Delta_{11})$ does not show any difference from predictions
of our null hypothesis $H_0$. For $\chi^2$ test the result is significant at
 $2\sigma$ level while for remaining four tests results are significant
on more than  $3\sigma$ level.
 
Our analysis leads to the conclusion that we observed significant alignment of
galaxies in our sample of rich Abell clusters. One should note that the
most powerful test is the Fourier test. It is not a surprise because during
previous analysis  of galactic alignment starting from \citet{h4} Fourier
test was the most sensitive one. Nearly the same significance level as
Fourier test shows the Kolmogorow test. One should note, that in the contrast
to analysis of individual structures where the autocorrelation test usually does not
lead to a significance conclusion (see for example \citep{g2,g3,g05}),
during our analysis the autocorrelation test is more powerful than the $\chi^2$ test.
 
\section{Conclusions}
 
We investigated statistical tests originally proposed by \citet{h4} for
analysis of the galactic orientations. Basing on analyzed  tests, the
method of analysis of the alignment of galaxies in clusters was proposed.
We analyzed the alignment of galaxies belonging to  247 Abell clusters
containing at least 100 members. The distributions of the position
angles for galaxies in each cluster were analyzed using statistical tests:
$\chi^2$ test, Fourier tests, Autocorrelation test and Kolmogorow test.
The mean value of the analyzed statistics was compared with theoretical
predictions as well as with results obtained from numerical simulations.
 
The statistical tests originally proposed by \citet{h4} for
analysis of the galactic orientations were the $\chi$ test, the Fourier test
and the autoccorelation test. We analyzed the autocorrelation test in more detail
and some improvements were proposed. It was shown that original \citet{h4}
result is an approximation which is not fully valid in our case.
We pointed out that the distribution of the number of galaxies with
orientations within the $k$-th  angular bin $N_k$ is in fact a polynomial
distribution and then in particular $N_K$ are not independent to each other.
In the result the expected value of $C$ statistics is equal $-1$ instead $0$ as
in original \citet{h4} paper. This difference is not significant in the
case of individual clusters because it is small with comparison to its standard
deviation $\sigma(C) \approx \sqrt{n}$. However in our case when we analyze
247 clusters and compute the average value of the statistics this difference begin
to be important. This is because variance of average values
$\sigma(\bar{C}) \approx \sqrt{n/247} = 0.3818$ starts to be significantly
smaller than a difference between our and approximated by Peebles expected
values of $C$. Separately we analyzed in detail the Fourier test.
We analyzed the properties of whole Fourier test as well as the
$\Delta_1/\sigma(\Delta_1)$, $\Delta/\sigma(\Delta)$ and $\Delta_{11}/\sigma(\Delta_{11})$
statistics. We compute the expected value and the variance of these statistics.
The results of our theoretical investigations were compared with the numerical
simulations.
 
Our analysis of the distributions of the position angles of galaxies in
rich Abell clusters shows  that the orientation of galaxies in  analyzed
cluster is not random i.e. we found an existence of alignment of galaxies
in the rich Abell galaxy clusters. Five statistical test show that
distribution of the position angles is not random at least on $3\,\sigma$
level. In all cases there was no significant difference when we analyzed
distribution of Equatorial position angles $p$ and Supergalactic position
angles $P$. Moreover $\Delta_{11}/\sigma(\Delta_{11})$ statistics do not
show any significant deviation from randomness. These two facts suggest that
observed alignment is not connected with equatorial plane (as expected) nor
with Supergalactic plane.
 
Our previous analysis \citep{g10,g11,f11} shows the dependency on the
alignment of galaxies in clusters and richness of the cluster which leads to
the conclusion  that the angular momentum of the cluster increases with the
mass of the structure. With such a dependency it is natural to expected that
in rich clusters significant alignment should be present. In the present
paper we confirmed this predictions.
 
Usually a dependence between the angular momentum and the mass of the
structure is presented as empirical relation $J\sim M^{5/3}$
\citep{Wesson79,Wesson83,Carrasco82,Brosche86}. In our opinion the observed
relation between the richness of the galaxy cluster and the alignment is due
to tidal torque, as suggested by \citet{HP88} and \citet{Catelan96}. Moreover, the
analysis of the linear tidal torque theory is pointing in the same direction
\citep{Noh06a,Noh06b}. They noticed the connection of the alignment with the
considered scale of the structure. However one should note that our result
is also compatible with the prediction of the Li model \citep{Li98,g03,g05}
in which galaxies form in the rotating universe.
 
In our further paper we would like to extend our consideration to analysis
of the distribution of two angles $\delta_D$ (the angle between the normal
to the galaxy plane and the main plane of the coordinate system), and $\eta$
(the angle between the projection of this normal onto the main plane and the
direction towards the zero initial meridian) describing the spatial
orientation of the galaxy plane. Moreover we would like to investigate if
effect found in the present paper depends on the cluster BM type.

\section*{Acknowledgments}
 
This research has made use of the NASA/IPAC Extragalactic Database (NED)
which is operated by the Jet Propulsion Laboratory, California Institute
of Technology, under contract with the National Aeronautics and Space
Administration. Author thanks Elena Panko and Piotr Flin for permission
to use unpublished data from their catalog and Stanis{\l}aw Jadach
for drawing my attention to RANLUX generator and for helpful remarks and
discussion on this problem. The author thanks anonymous referee for detailed 
remarks which helped to improve the original manuscript.

\clearpage
 
\begin{figure}
\includegraphics[angle=270,scale=0.30]{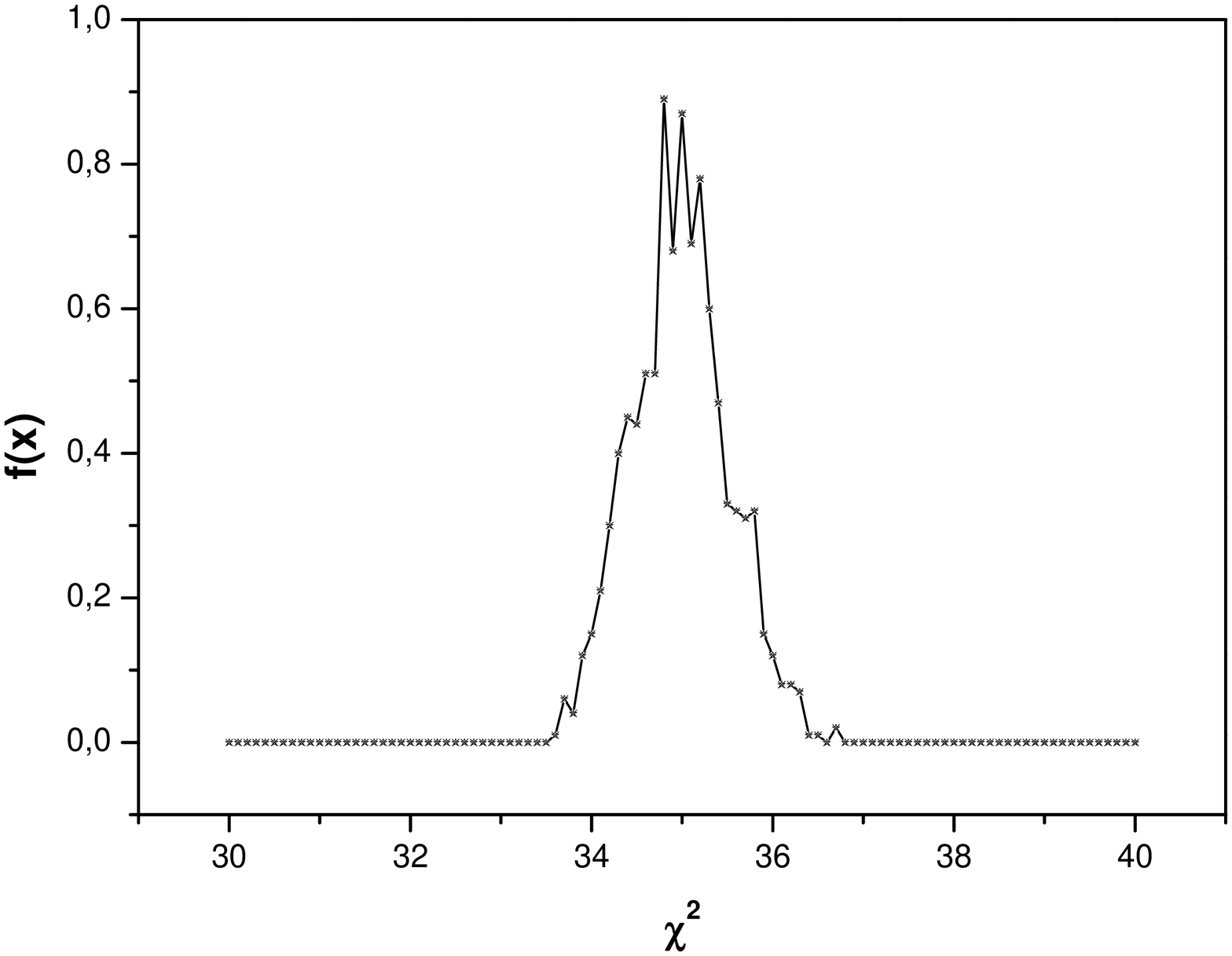}
\includegraphics[angle=270,scale=0.30]{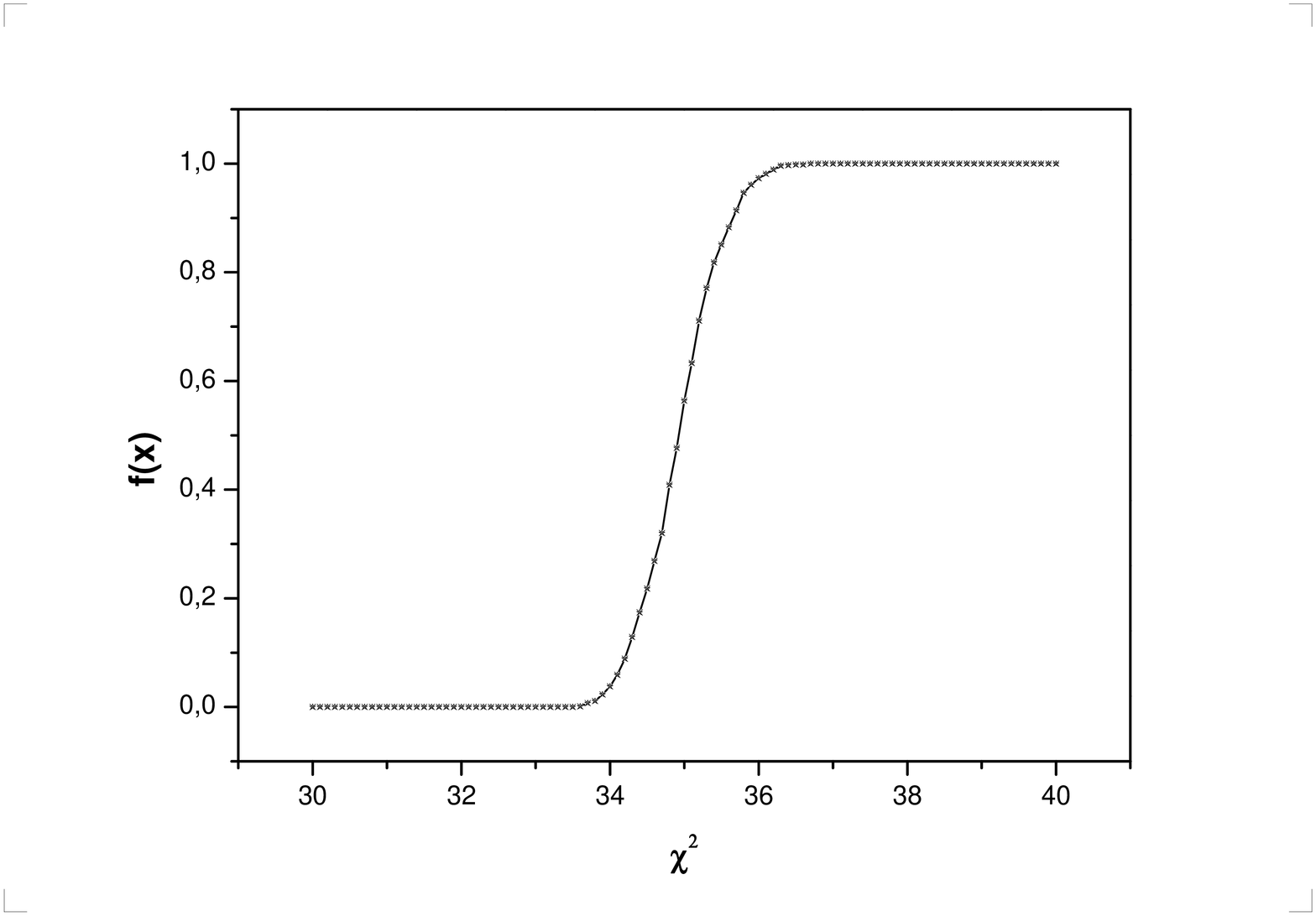}\\
\includegraphics[angle=270,scale=0.30]{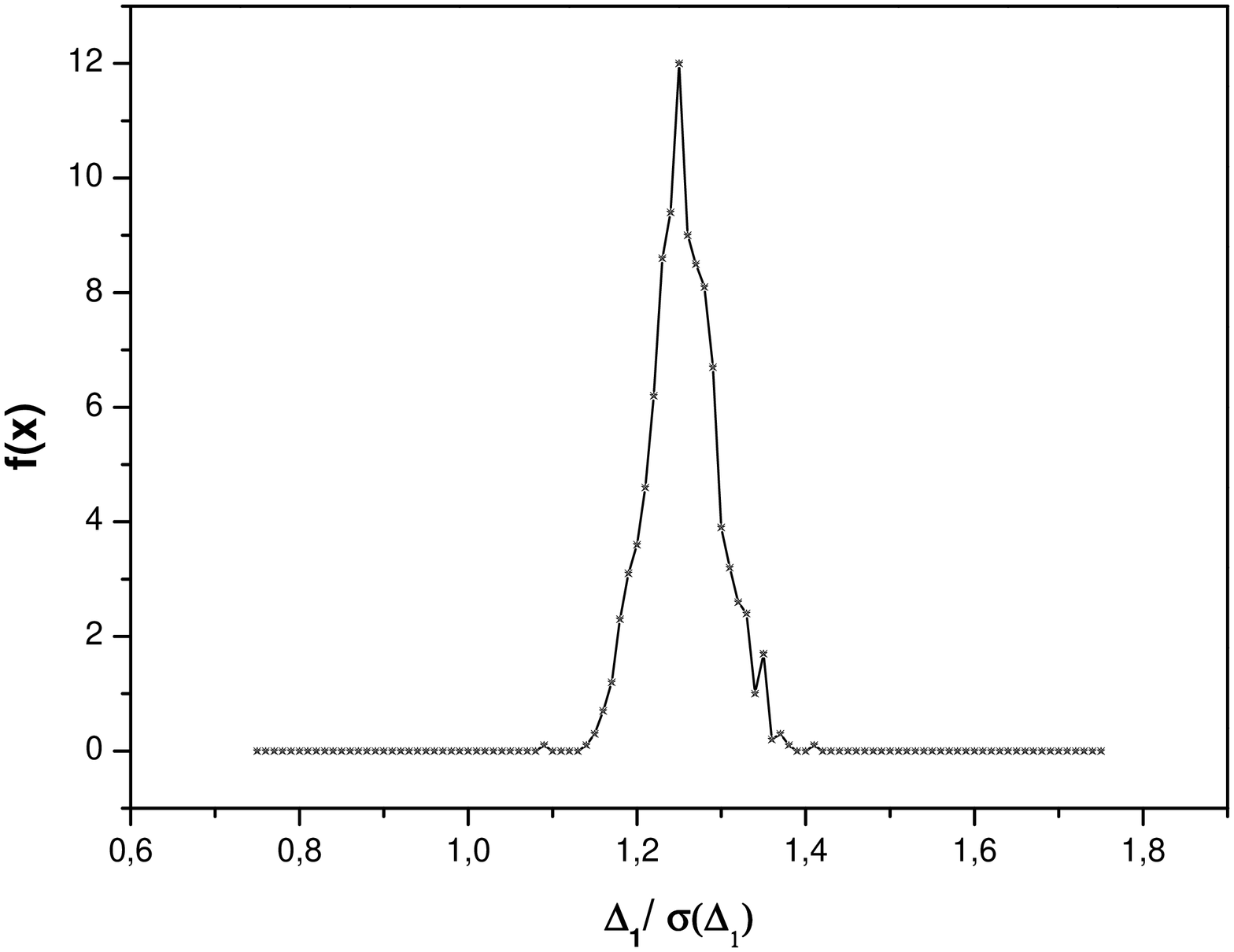}
\includegraphics[angle=270,scale=0.30]{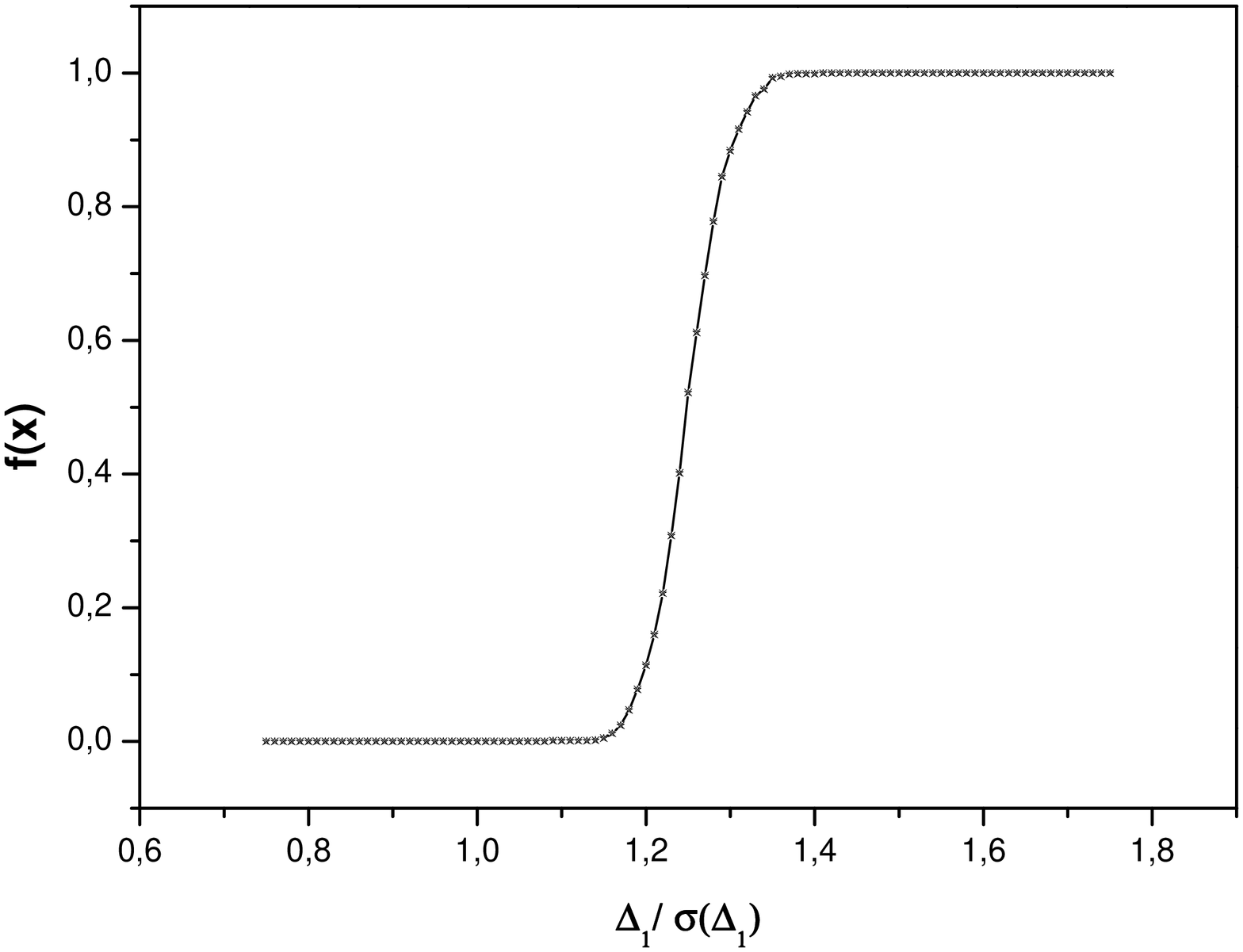}\\
\includegraphics[angle=270,scale=0.30]{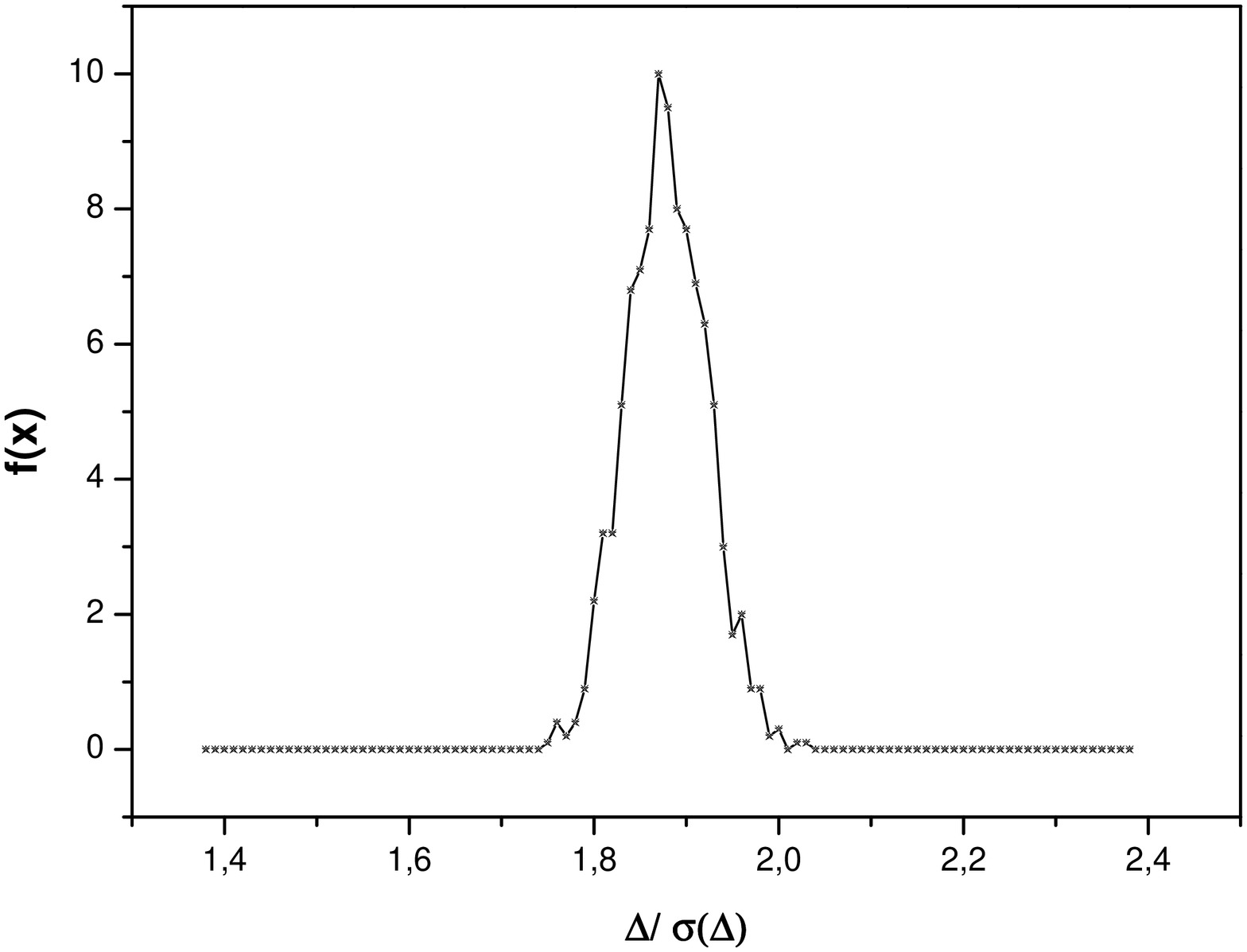}
\includegraphics[angle=270,scale=0.30]{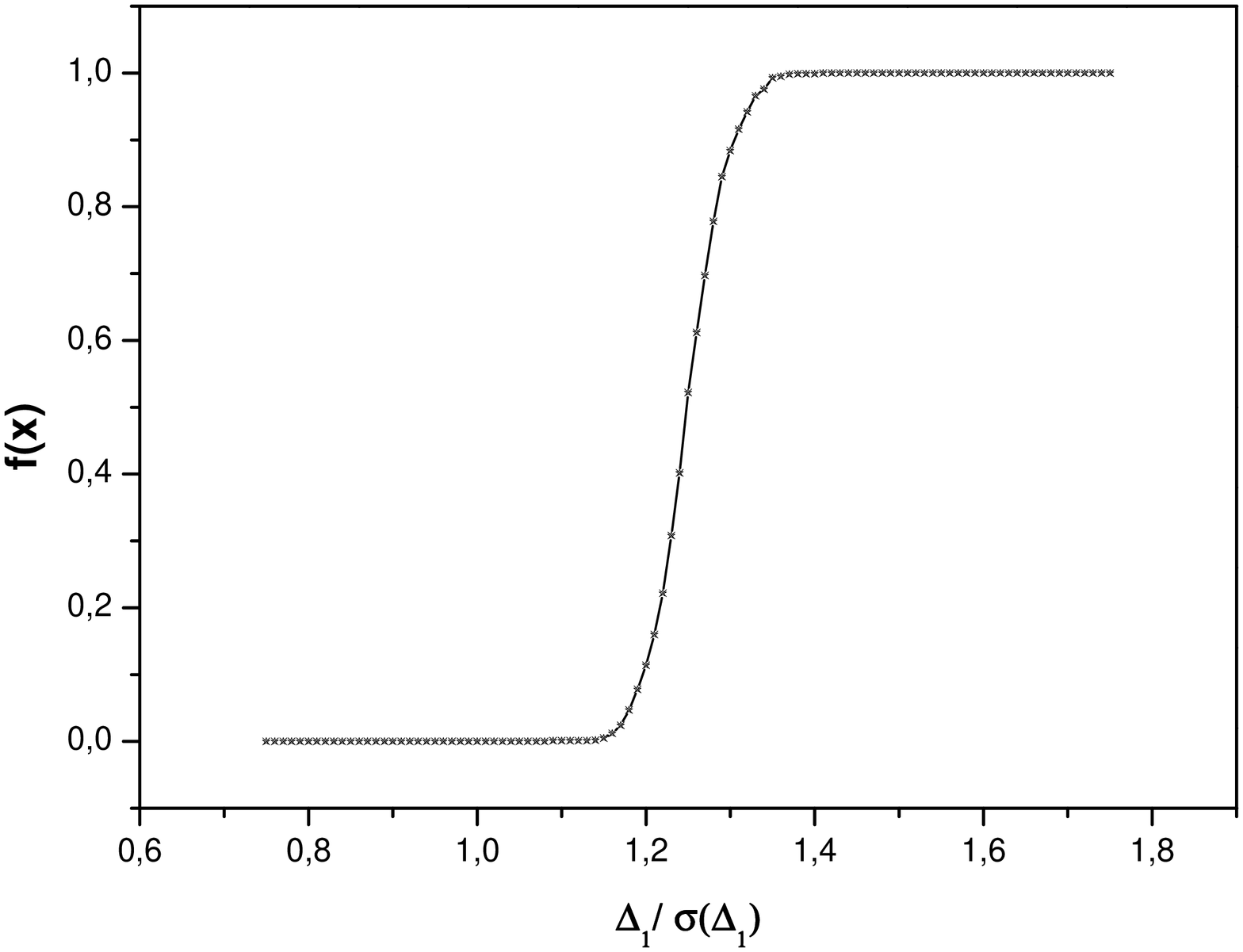}\\
\caption{The Probability Density Function (PDF) (left panel) and
Cumulative Distribution Function (CDF) (right panel) for analyzed statistics.
The figure was obtained from 1000 simulations of sample of 247 cluster
each with number of members galaxies the same as in the real cluster.
From up to down we present statistics: 
$\chi^2$, $\Delta_{1}/\sigma(\Delta_{1})$, $\Delta/\sigma(\Delta)$.
\label{fig:f1}}
\end{figure}
 
\clearpage
 
\begin{figure}
\includegraphics[angle=270,scale=0.30]{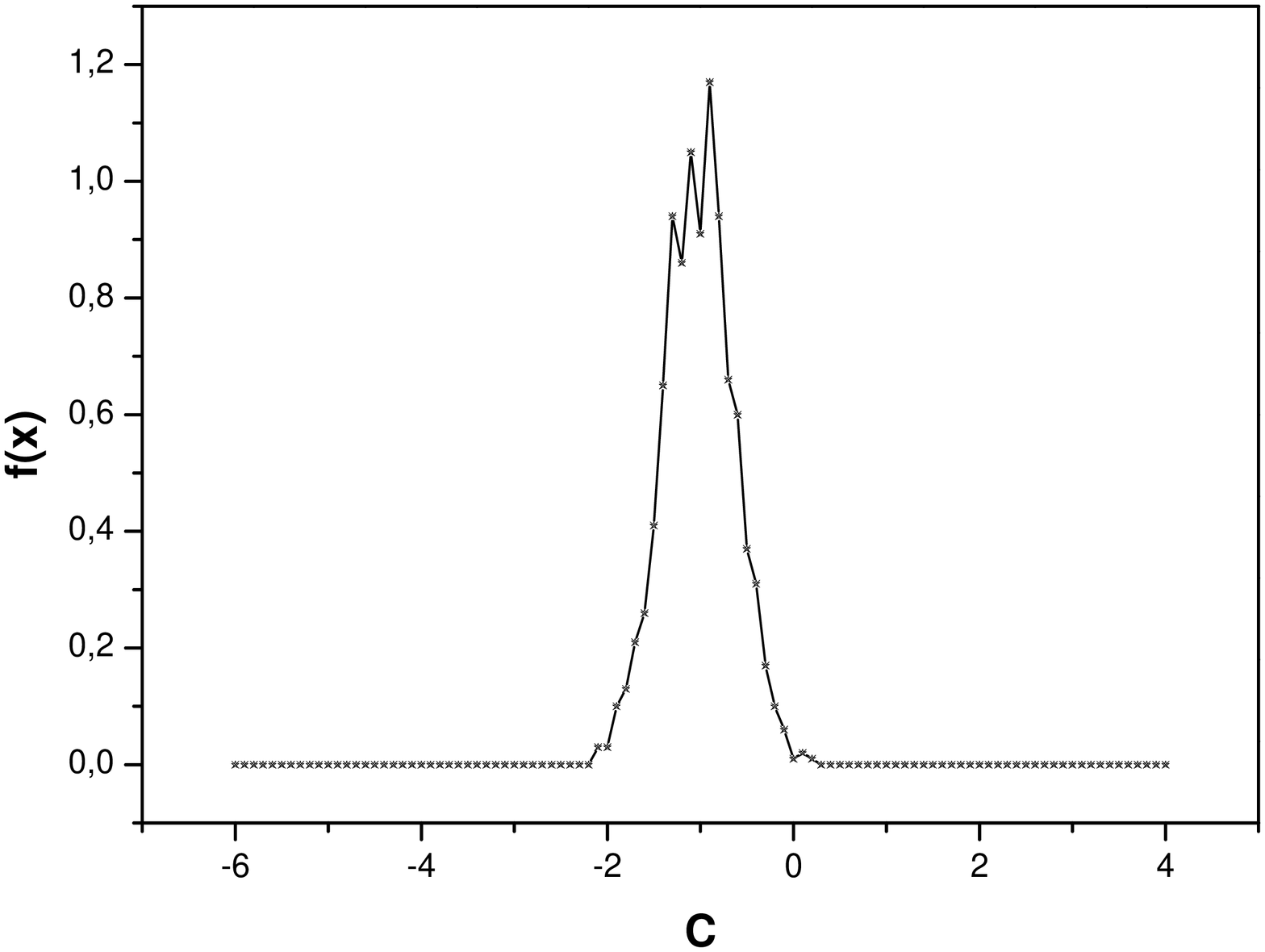}
\includegraphics[angle=270,scale=0.30]{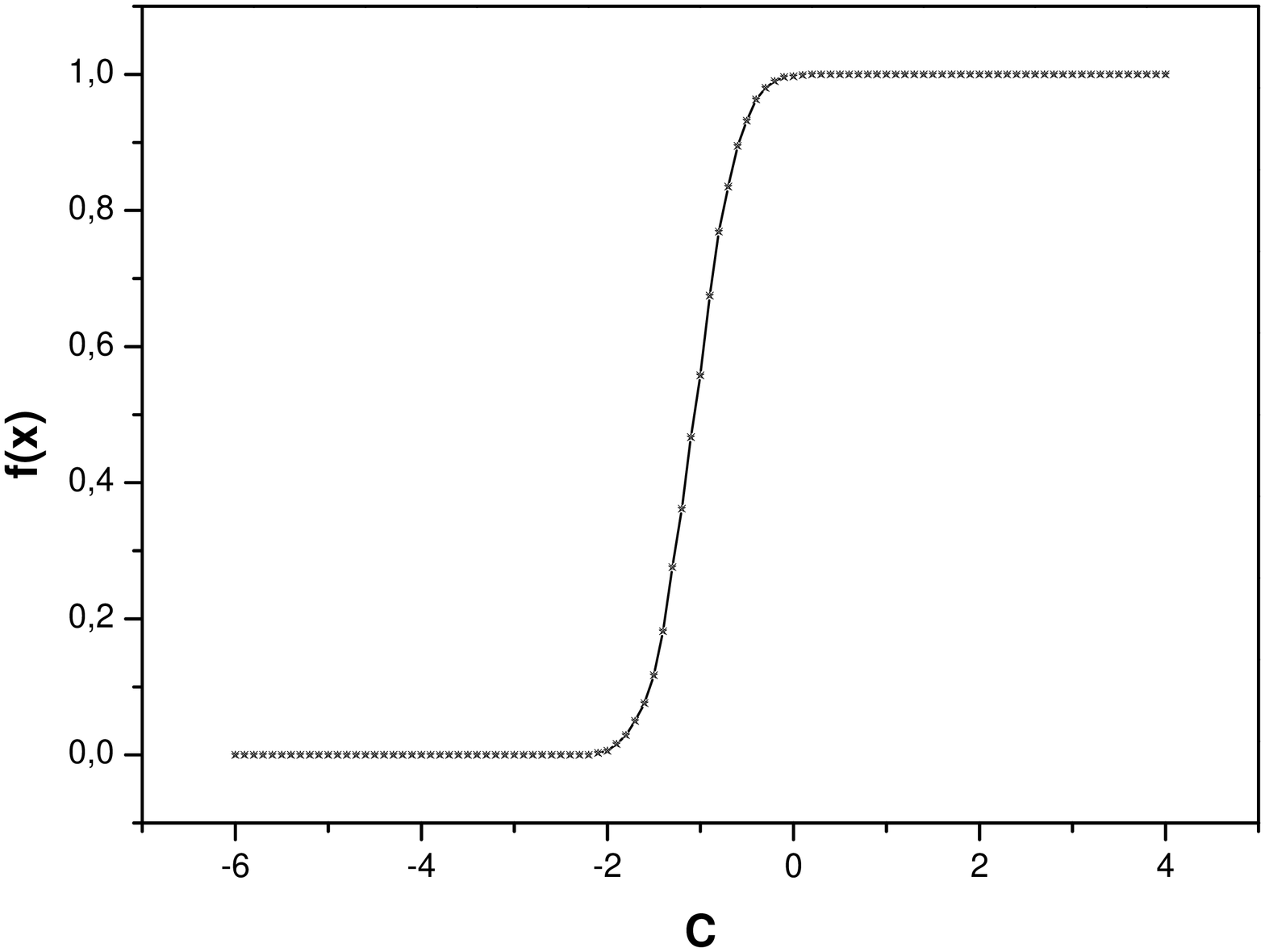}\\
\includegraphics[angle=270,scale=0.30]{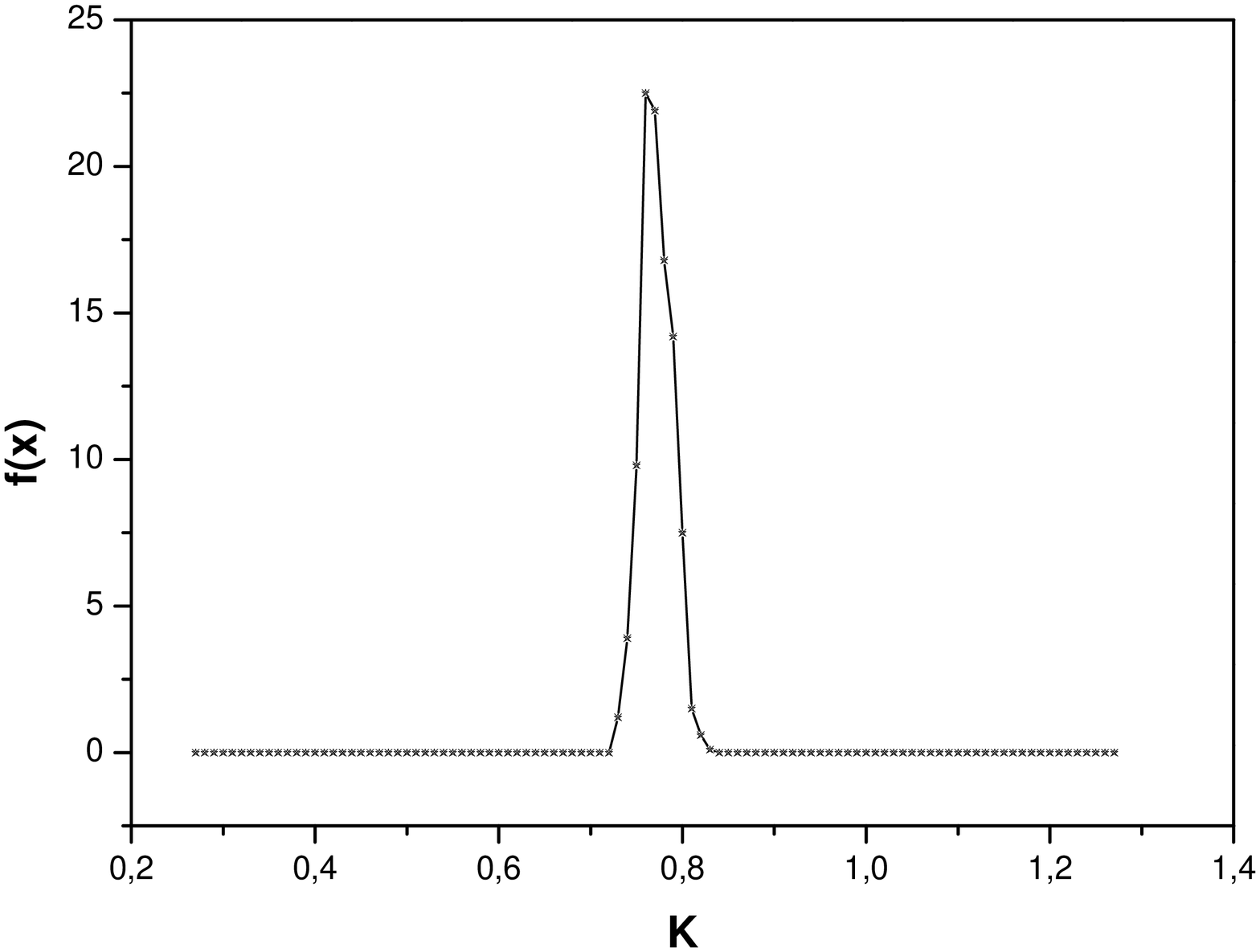}
\includegraphics[angle=270,scale=0.30]{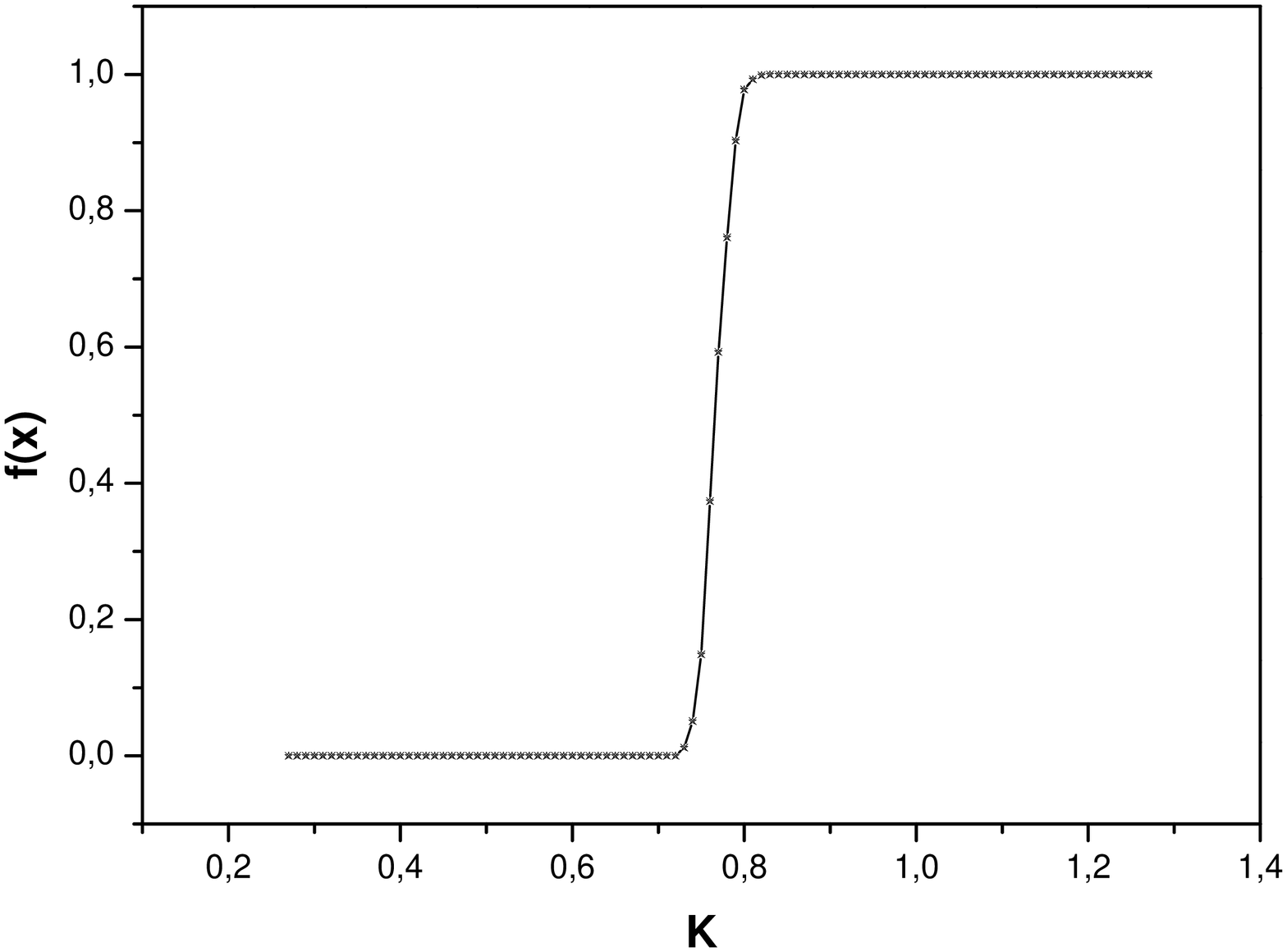}\\
\includegraphics[angle=270,scale=0.30]{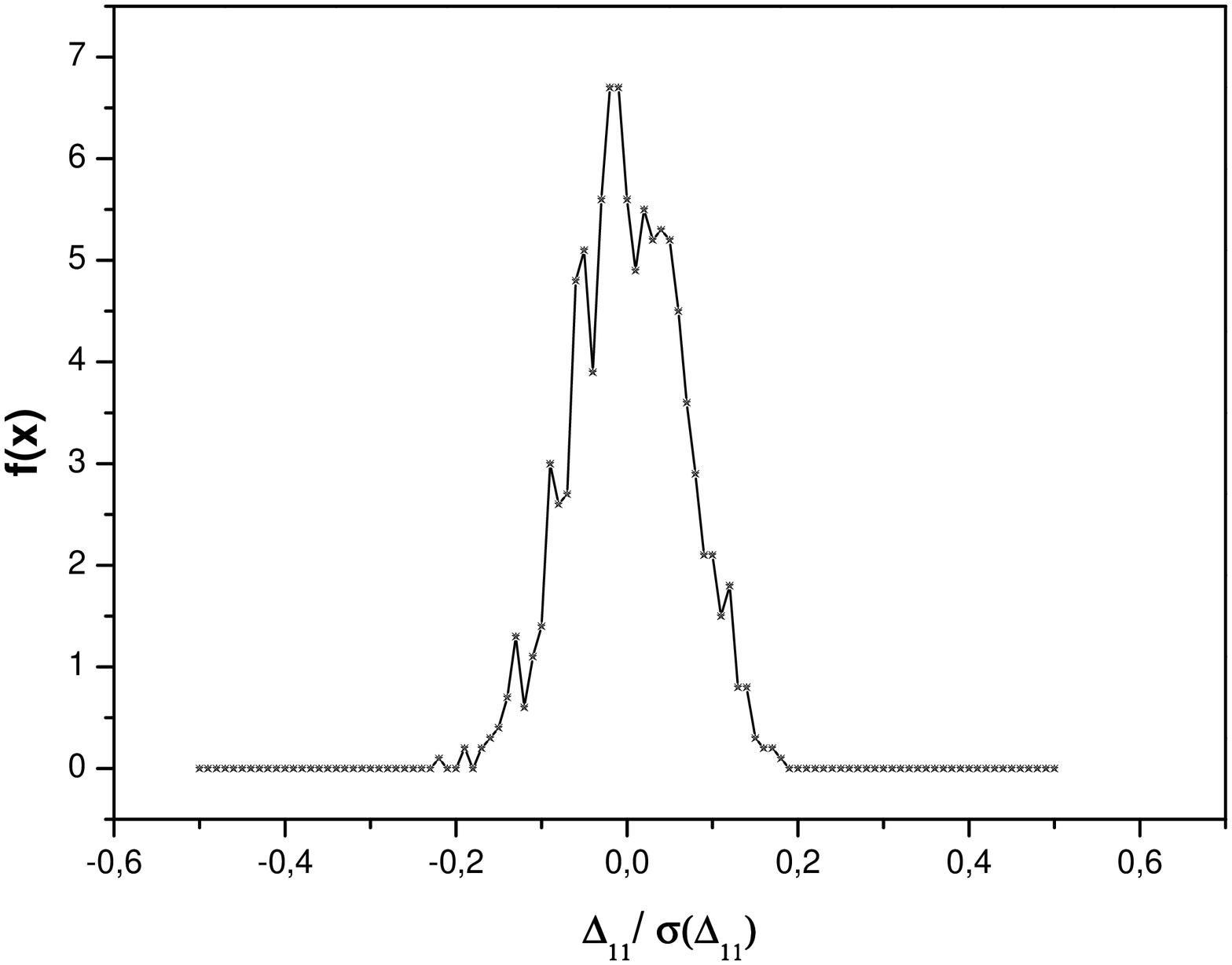}
\includegraphics[angle=270,scale=0.30]{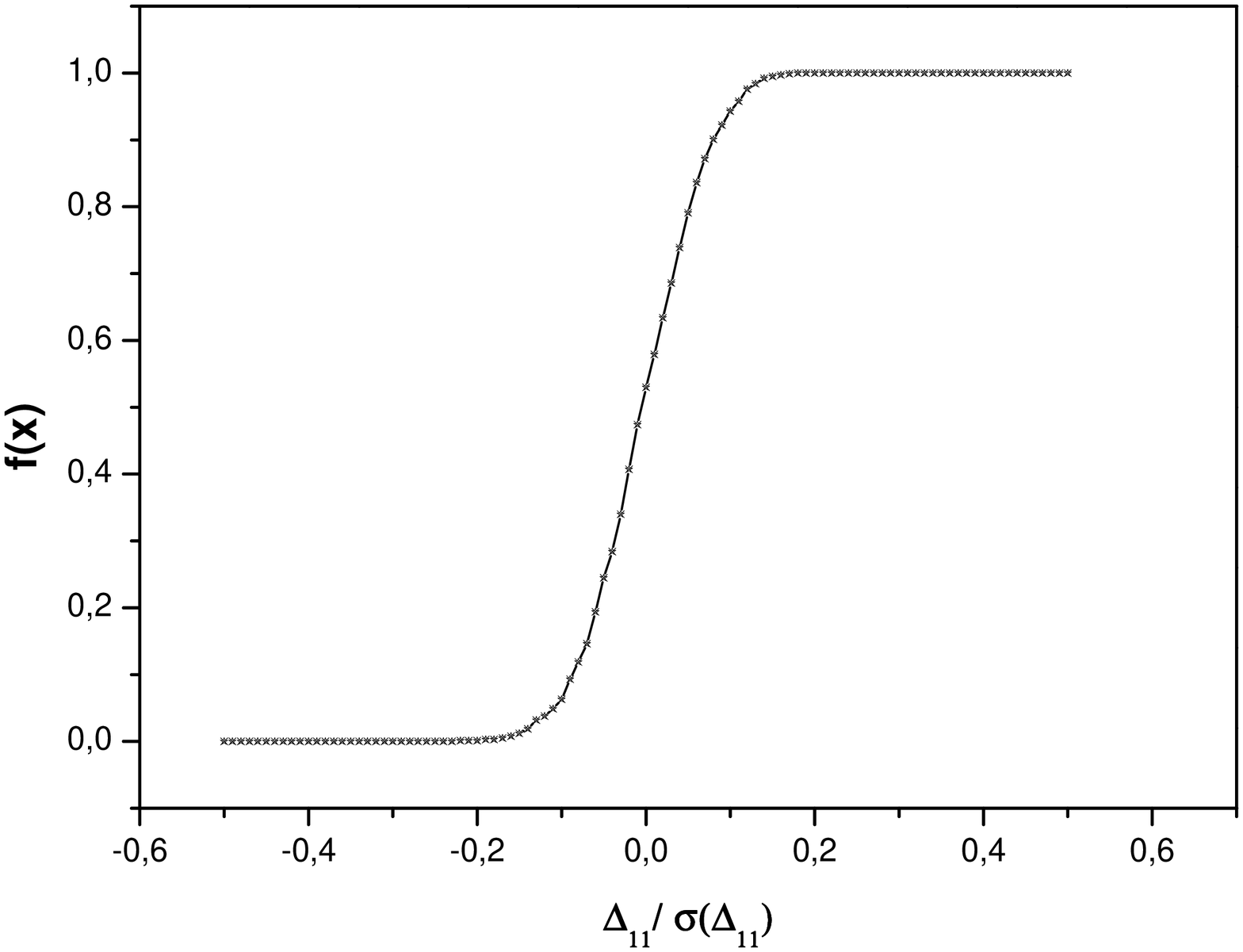}\\
\caption{The Probability Density Function (PDF) (left panel) and
Cumulative Distribution Function (CDF) (right panel) for analyzed statistics.
The figure was obtained from 1000 simulations of sample of 247 cluster
each with number of members galaxies the same as in the real cluster.
From up to down we present statistics:
$C$, $\lambda$, $\Delta_{11}/\sigma(\Delta_{11})$.
\label{fig:f2}}
\end{figure}
 
\clearpage 
 
\begin{table}
\begin{center}
\caption{The comparision of different numerical generators}
\label{tab:t1}
\begin{tabular}{cccccc}
\tableline \tableline
Generator&Test&$\bar{x}$&$\sigma(x)$&$\sigma(\bar{x})$&$\sigma(\sigma(x))$\\
\tableline
$Lahey$    &$\chi^2$                         &$34.9741$&$0.5271$&$0.0166$&$0.0117$\\
           &$C$                              &$-0.9981$&$0.3743$&$0.0118$&$0.0083$\\
           &$\Delta_{11}/\sigma(\Delta_{11})$&$-0.0040$&$0.0616$&$0.0019$&$0.0014$\\
\tableline
$Ran1_{pt}$&$\chi^2$                         &$34.4993$&$0.5157$&$0.0163$&$0.0115$\\
           &$C$                              &$-0.5145$&$0.3569$&$0.0113$&$0.0080$\\
           &$\Delta_{11}/\sigma(\Delta_{11})$&$ 0.0018$&$0.0613$&$0.0019$&$0.0014$\\
\tableline
$Ran1_{nr}$&$\chi^2$                         &$34.9604$&$0.5320$&$0.0168$&$0.0119$\\
           &$C$                              &$-1.0018$&$0.3836$&$0.0121$&$0.0086$\\
           &$\Delta_{11}/\sigma(\Delta_{11})$&$ 0.0282$&$0.0632$&$0.0020$&$0.0014$\\
\tableline
$GGUBS$    &$\chi^2$                         &$34.9874$&$0.5505$&$0.0174$&$0.0123$\\
           &$C$                              &$-1.0110$&$0.3765$&$0.0119$&$0.0084$\\
           &$\Delta_{11}/\sigma(\Delta_{11})$&$ 0.0021$&$0.0637$&$0.0020$&$0.0014$\\
\tableline
$RANLUX$   &$\chi^2$                         &$34.9978$&$0.5442$&$0.0172$&$0.0121$\\
           &$C$                              &$-0.9917$&$0.3899$&$0.0123$&$0.0087$\\
           &$\Delta_{11}/\sigma(\Delta_{11})$&$-0.0010$&$0.0643$&$0.0020$&$0.0014$\\
\tableline
\end{tabular}
\end{center}
\end{table}
 
\clearpage
 
\begin{table}
\begin{center}
\caption{The result of numerical simulation - sample of 247 cluster
each with 2360 galaxies}
\label{tab:t2}
\begin{tabular}{ccccc}
\tableline \tableline
Test&$\bar{x}$&$\sigma(x)$&$\sigma(\bar{x})$&$\sigma(\sigma(x))$\\
\tableline
$\chi^2$                         &$34.9978$&$0.5442$&$0.0172$&$0.0121$\\
$\Delta_{1}/\sigma(\Delta_{1})$  &$ 1.2524$&$0.0424$&$0.0013$&$0.0009$\\
$\Delta/\sigma(\Delta)$          &$ 1.8794$&$0.0460$&$0.0014$&$0.0010$\\
$C$                              &$-0.9917$&$0.3899$&$0.0123$&$0.0087$\\
$\lambda$                        &$ 0.7708$&$0.0166$&$0.0005$&$0.0003$\\
$\Delta_{11}/\sigma(\Delta_{11})$&$-0.0010$&$0.0643$&$0.0020$&$0.0014$\\
\tableline
\end{tabular}
\end{center}
\end{table}
 
\clearpage
 
\begin{table}
\begin{center}
\caption{The result of numerical simulation - sample of 247 cluster
each with number of members galaxies the same as in the real cluster.}
\label{tab:t3}
\begin{tabular}{ccccc}
\tableline \tableline
Test&$\bar{x}$&$\sigma(x)$&$\sigma(\bar{x})$&$\sigma(\sigma(x))$\\
\tableline
$\chi^2$                         &$34.9798$&$0.5364$&$0.0170$&$0.0120$\\
$\Delta_{1}/\sigma(\Delta_{1})$  &$ 1.2550$&$0.0419$&$0.0013$&$0.0009$\\
$\Delta/\sigma(\Delta)$          &$ 1.8788$&$0.0436$&$0.0014$&$0.0010$\\
$C$                              &$-1.0195$&$0.3749$&$0.0119$&$0.0084$\\
$\lambda$                        &$ 0.7720$&$0.0168$&$0.0005$&$0.0004$\\
$\Delta_{11}/\sigma(\Delta_{11})$&$ 0.0014$&$0.0645$&$0.0020$&$0.0014$\\
\tableline
\end{tabular}
\end{center}
\end{table}
 
\clearpage
 
\begin{table}
\begin{center}
\caption{The result of numerical simulation - sample of 247 cluster
each with number of members galaxies the same as in the real cluster
but only galaxies brighter than $m_3+3$ are taken into account.}
\label{tab:t4}
\begin{tabular}{ccccc}
\tableline \tableline
Test&$\bar{x}$&$\sigma(x)$&$\sigma(\bar{x})$&$\sigma(\sigma(x))$\\
\tableline
$\chi^2$                         &$34.9902$&$0.5309$&$0.0168$&$0.0119$\\
$\Delta_{1}/\sigma(\Delta_{1})$  &$ 1.2564$&$0.0418$&$0.0013$&$0.0009$\\
$\Delta/\sigma(\Delta)$          &$ 1.8820$&$0.0434$&$0.0014$&$0.0010$\\
$C$                              &$-1.0007$&$0.3716$&$0.0118$&$0.0083$\\
$\lambda$                        &$ 0.7724$&$0.0163$&$0.0005$&$0.0004$\\
$\Delta_{11}/\sigma(\Delta_{11})$&$-0.0006$&$0.0636$&$0.0020$&$0.0014$\\
\tableline
\end{tabular}
\end{center}
\end{table}
 
\clearpage
 
\begin{table}
\begin{center}
\caption {The value of analyzed statistics real sample of 247 Abell clusters.}
\label{tab:t5}
\begin{tabular}{c|c|cc|cc}
\tableline \tableline
\multicolumn{1}{c}{}&
\multicolumn{1}{c}{}&
\multicolumn{2}{c}{Equatorial coordinates}&
\multicolumn{2}{c}{Supergalactic coordinates}\\
\tableline
Sample&Test&$\bar{x}$&$\sigma(\bar{x})$&$\bar{x}$&$\sigma(\bar{x})$\\
\tableline
A&$\chi^2$                         &$36.8591$&$0.5924$&$36.7899$&$0.6315$\\
 &$\Delta_{1}/\sigma(\Delta_{1})$  &$ 1.7046$&$0.0622$&$ 1.7021$&$0.0626$\\
 &$\Delta/\sigma(\Delta)$          &$ 2.2663$&$0.0594$&$ 2.2746$&$0.0591$\\
 &$C$                              &$ 1.1940$&$0.4530$&$ 1.1220$&$0.4237$\\
 &$\lambda$                        &$ 0.9177$&$0.0240$&$ 0.9138$&$0.0220$\\
 &$\Delta_{11}/\sigma(\Delta_{11})$&$-0.0005$&$0.0855$&$ 0.0940$&$0.0924$\\
\tableline
B&$\chi^2$                         &$36.4000$&$0.6072$&$36.2919$&$0.6124$\\
 &$\Delta_{1}/\sigma(\Delta_{1})$  &$ 1.6283$&$0.0577$&$ 1.6316$&$0.0578$\\
 &$\Delta/\sigma(\Delta)$          &$ 2.2055$&$0.0565$&$ 2.2199$&$0.0554$\\
 &$C$                              &$ 0.8843$&$0.4355$&$ 0.7863$&$0.4212$\\
 &$\lambda$                        &$ 0.8928$&$0.0224$&$ 0.8934$&$0.0210$\\
 &$\Delta_{11}/\sigma(\Delta_{11})$&$ 0.0023$&$0.0826$&$ 0.0810$&$0.0866$\\
\tableline
\end{tabular}
\end{center}
\end{table}
 
\end{document}